\documentclass[journal]{IEEEtran}

\usepackage{amsfonts}
\usepackage{amsmath}
\usepackage{amssymb}
\usepackage{graphicx}
\usepackage{mathtools}
\usepackage{adjustbox}
\usepackage{hyperref}
\usepackage{comment}
\usepackage{cite} %  if labels are of an ordered list, reduce with a hyphen

\usepackage[caption=false, font=footnotesize]{subfig} % for subfigures

\usepackage{multirow} % Add this to your LaTeX preamble if not already included
\usepackage{nicematrix}

\usepackage{amsthm} % for the proof

\usepackage{soul}  % for highlighting a text

\usepackage[normalem]{ulem}

\newtheorem{example}{Example}
\newtheorem{remark}{Remark}

\usepackage{array} % for more control over columns
\usepackage{lipsum} % just for sample text

\usepackage[makeroom]{cancel} % To cross something

\usepackage{xcolor} % text color
\newcommand{\B}[1]{\textcolor{blue}{#1}}
\newcommand{\R}[1]{\textcolor{red}{#1}}
\newcommand{\G}[1]{\textcolor{green}{#1}}
\newcommand{\PP}[1]{\textcolor{purple}{#1}}

\usepackage{tikz}
\usetikzlibrary{intersections}

\usetikzlibrary{circuits.logic.US, positioning}
\usetikzlibrary{shapes.geometric, arrows}

\usepackage{algorithm}
\usepackage{algpseudocode}

\usepackage[a-1b]{pdfx}

\IEEEoverridecommandlockouts % to add the footnoote at the first page.

\begin{document}

\title{On the High-Rate FDPC Codes: Construction, Encoding, and a Generalization}

\author{Mohsen~Moradi\textsuperscript{\href{https://orcid.org/0000-0001-7026-0682}{\includegraphics[scale=0.06]{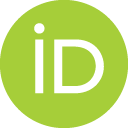}}},
Sheida~Rabeti\textsuperscript{\href{https://orcid.org/0009-0005-5040-4542}{\includegraphics[scale=0.06]{figs/ORCID}}},
and Hessam~Mahdavifar\textsuperscript{\href{https://orcid.org/0000-0001-9021-1992}{\includegraphics[scale=0.06]{figs/ORCID}}}\\
\thanks{The authors are with the Department of Electrical \& Computer Engineering, Northeastern University, Boston MA-02115, USA (e-mail: m.moradi@northeastern.edu, rabeti.s@northeastern.edu, h.mahdavifar@northeastern.edu).}
\thanks{This work was supported by the National Science Foundation (NSF) under Grant CCF-2415440 and the Center for Ubiquitous Connectivity (CUbiC) under the JUMP 2.0 program.}
}

% \author{Mohsen~Moradi,
% Sheida~Rabeti,
% and Hessam~Mahdavifar\\
% \IEEEauthorblockA{Department of Electrical \& Computer Engineering, Northeastern University, Boston MA 02115, USA\\
% Email: \{m.moradi, rabeti.s, h.mahdavifar\}@northeastern.edu}
% \thanks{This work was supported by the National Science Foundation (NSF) under Grant CCF-2415440 and the Center for Ubiquitous Connectivity (CUbiC) under the JUMP 2.0 program.}
% }

\maketitle
\begin{abstract} 
Recently introduced Fair-Density Parity-Check (FDPC) codes, targeting high-rate applications, offer superior error-correction performance (ECP) compared to 5G Low-Density Parity-Check (LDPC) codes, given the same number of message-passing decoding iterations. 
In this paper, we present a novel construction method for FDPC codes, introduce a generalization of these codes, and propose a low-complexity encoding algorithm. Numerical results demonstrate the fast convergence of the message-passing decoder for FDPC codes.
\end{abstract}
\begin{IEEEkeywords}
FDPC codes, LDPC codes, encoding, minimum distance, channel coding, belief propagation, normalized min-sum.
\vspace{-7pt}
\end{IEEEkeywords}

%##########################################################################################################
\vspace{-7pt}
\section{Introduction}

\IEEEPARstart{T}{he} Low-Density Parity-Check (LDPC) codes \cite{gallager1962low, mackay1995good} can approach channel capacity for large block lengths when decoded using message-passing algorithms such as belief propagation (BP) or simplified variants like the normalized min-sum algorithm. 
These message-passing algorithms have a complexity that scales linearly with the code block length. 
LDPC codes have been adopted for data channels in 5G communication systems \cite{3GPP_2018} and are considered one of the leading candidates for next-generation 6G communication systems \cite{wang2023road}.
Some designs of LDPC codes can be found in \cite{mitchell2015spatially, tanner2004ldpc, pusane2011deriving, mitchell2014quasi, costello2014spatially, bocharova2022design}.

Polar codes, introduced by Ar{\i}kan \cite{arikan2009channel}, have also been adopted for control channels in 5G communication and are similarly regarded as strong candidates for 6G communication systems.
Polar codes are the first class of codes that are theoretically proven to be able to achieve symmetric channel capacity with explicit constructions and low-complexity encoding and decoding.
The capacity-achieving performance of polar codes is realized with successive cancellation (SC) decoding in an asymptotic sense. 
At short or moderate block lengths, the performance of SC decoding falls significantly short of the theoretical limits. 
To enhance SC decoder performance, cyclic-redundancy check (CRC)-aided SC list (CA-SCL) decoding has been proposed, which significantly improves the performance of polar codes at short and moderate lengths, surpassing the performance of LDPC codes \cite{tal2015list}.

The newly introduced Fair-Density Parity-Check (FDPC) codes, designed for high-rate applications, demonstrate superior error-correction performance (ECP) compared to 5G-LDPC codes with the same number of message-passing algorithm iterations \cite{mahdavifar2024high}. The FDPC code construction consists of a carefully designed base parity-check matrix with desired combinatorial properties together with random permutations and specific cascading methods for general constructions in the high-rate regime. 

In this paper, we propose a method for constructing the base matrix of an FDPC code meeting a given minimum distance, and we demonstrate its effectiveness using two example base matrices.
The first base matrix is derived by rearranging the columns of the base matrix introduced in \cite{mahdavifar2024high}, with certain modifications. 
We present a low-complexity encoding algorithm and discuss the ECP of FDPC codes after puncturing.
Our designed base matrices consist of a collection of bidiagonal matrices. We utilize a bidiagonal matrix for the encoding of our proposed FDPC codes, which is similar to how the encoding is done for the 5G-LDPC codes \cite{richardson2018design}. Bidiagonal matrices can also be observed in an equivalent construction of polarization-adjusted convolutional codes \cite{moradi2024polarization}.
For $(256, 164)$ codes, our proposed FDPC construction, using only 5 iterations of the normalized min-sum algorithm, achieves approximately a $0.5$~dB coding gain over the 5G LDPC code decoded with the same algorithm and the polar code decoded with belief propagation (BP) \cite{arikan2010polar}, both using 50 iterations at a frame error rate (FER) level of $10^{-3}$. 
It also provides about a $1$~dB gain compared to the polar code decoded with 5 iterations.
Furthermore, BP decoding of polar codes exhibits latency that is $\mathcal{O}(\log_2(N))$ times higher than that of the FDPC decoder. 
For $(1024, 844)$ codes, our proposed FDPC code, with only 12 decoding iterations, achieves an FER performance comparable to that of the CA-SCL decoding of 5G polar codes with a list size of 8, as well as the normalized min-sum decoding of 5G LDPC codes with 50 iterations. Notably, CA-SCL decoding incurs significantly higher latency compared to the message-passing algorithms used for FDPC and LDPC codes.

The remainder of the paper is organized as follows.
% In Section \ref{sec:LDPC}, we provide an explanation of Gallager's LDPC code construction.
Section \ref{sec:base1} introduces our first proposed construction for FDPC codes and describes the corresponding encoder.
In Section \ref{sec:base2}, we extend and generalize the initial construction.
Section \ref{sec:numerical} presents a comparison between our proposed construction and various other codes.
Finally, Section \ref{sec:Conclusion} summarizes the key contributions and concludes the paper.

% \section{LDPC Code Example}\label{sec:LDPC}

% LDPC codes have a very sparse parity check matrix.
% Gallager introduced regular LDPC codes that have a $w_c$ number of ones in each column and $w_r$ number of ones in each row.
% A toy example given by Gallager with $w_r = 4$ is the parity check matrix that has the base matrix

% \[
% \scriptsize
% \setlength{\arraycolsep}{3pt} % Adjust column spacing
% \mathbf{G}_0 =
% \left[
% \begin{array}{cccccccccccccccc}
% 1 & 1 & 1 & 1 & 0 & 0 & 0 & 0 & 0 & 0 & 0 & 0 & 0 & 0 & 0 & 0 \\
% 0 & 0 & 0 & 0 & 1 & 1 & 1 & 1 & 0 & 0 & 0 & 0 & 0 & 0 & 0 & 0 \\
% 0 & 0 & 0 & 0 & 0 & 0 & 0 & 0 & 1 & 1 & 1 & 1 & 0 & 0 & 0 & 0 \\
% 0 & 0 & 0 & 0 & 0 & 0 & 0 & 0 & 0 & 0 & 0 & 0 & 1 & 1 & 1 & 1 \\
% \end{array}
% \right].
% \]
% To have $w_c = 3$, one can cascade the matrix $\mathbf{G}_0$ with $\pi_1(\mathbf{G}_0)$ and $\pi_2(\mathbf{G}_0)$ to obtain
% \[
% \mathbf{H} = 
% \left[
% \begin{array}{c}
%       \mathbf{G}_0 \\
%      \pi_1(\mathbf{G}_0) \\
%      \pi_2(\mathbf{G}_0)
% \end{array}
% \right],
% \]
% where $\pi_1$ and $\pi_2$ randomly permute the columns of the base matrix $\mathbf{G}_0$.

\section{FDPC Code Construction base-1}\label{sec:base1}

In this section, we present the construction of a base matrix \( \textbf{H}_{\text{base-(t,1)}} \) with \( 2t \) rows and $t^2$ columns, for a $t \in {\mathbb{N}}$, where each column contains exactly two ones with the rest of entries set to zero. 
In this matrix, the first \( 2t-1 \) columns have two ones with no zeros between them. 
The next \( 2t-3 \) columns have two zeros between the ones. 
The next \( 2t-5 \) columns have four zeros between the ones. This continues in the same fashion with the last column having $2t-2$ zeros between the two ones placed in the first and last rows. This base matrix is equivalent to the one introduced in \cite{mahdavifar2024high}, with the columns rearranged.

\begin{figure}[htbp]
\centering
	\subfloat[\label{fig: decagon_loop4} Minimum cycle of length 4.]{
		\includegraphics[width=0.45\linewidth]{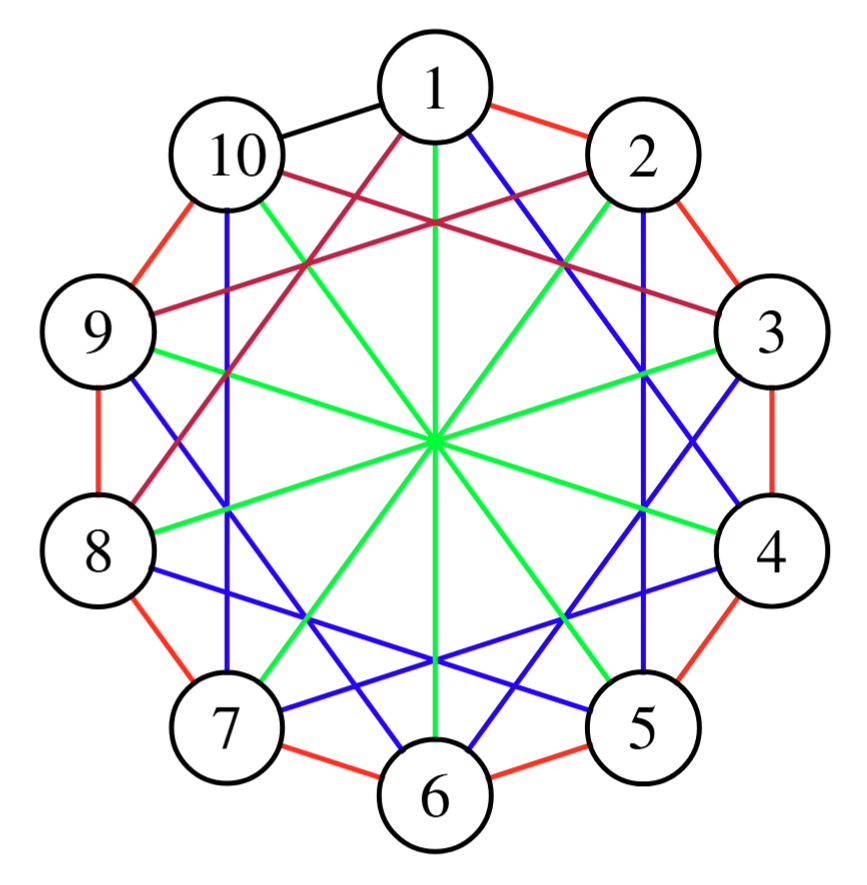}
	} \hspace{-0.04\linewidth}
	\subfloat[\label{fig: decagon_loop6} Minimum cycle of length 6.]{
		\includegraphics[width=0.47\linewidth]{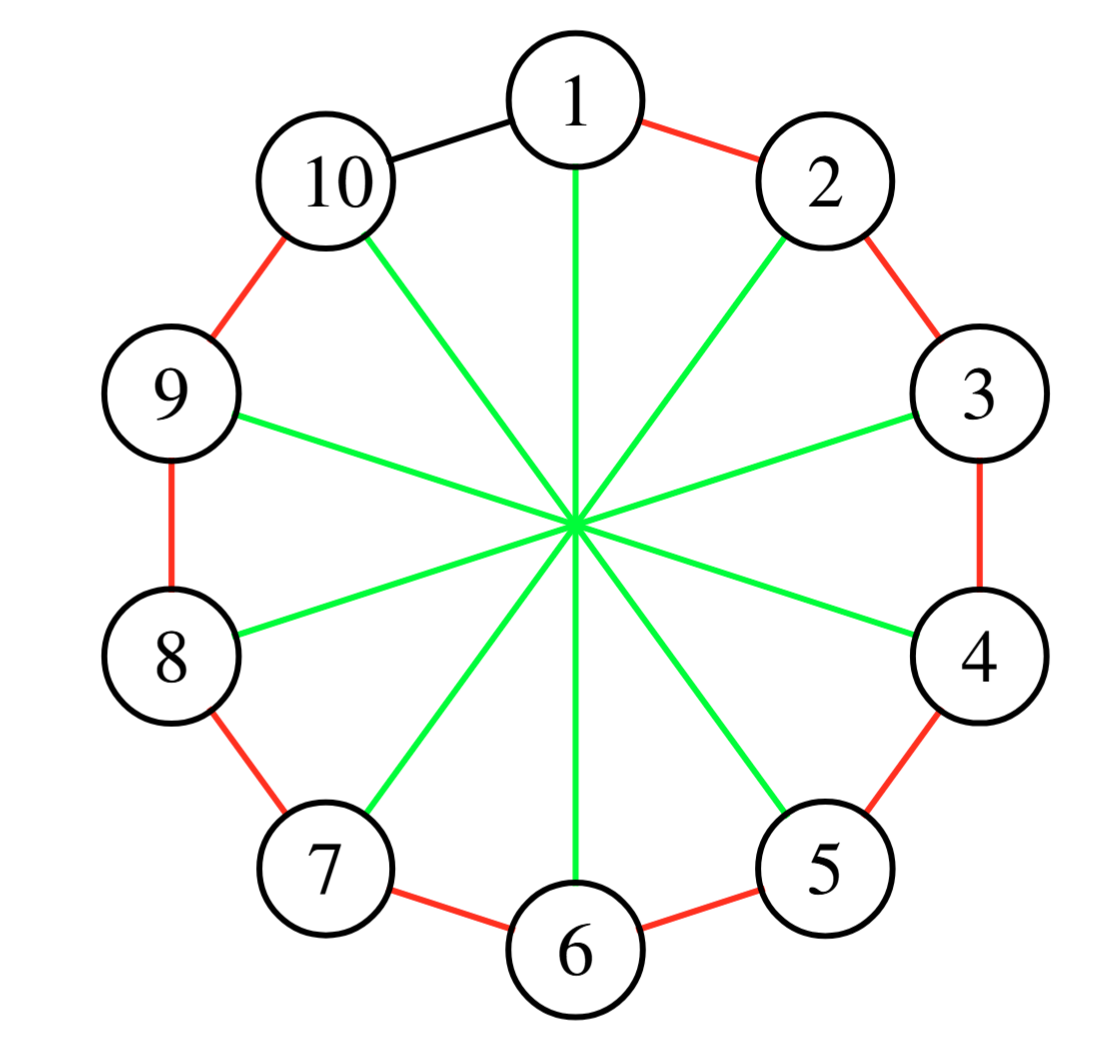}
	}
	\caption{Decagon graph representation of the base matrix structure for $t=5$: (a) with a minimum cycle of length 4, (b) with a minimum cycle of length 6.}
	\label{fig:decagon_combined}
\end{figure}

\begin{example}

For \( t = 5 \), the base matrix \( \textbf{H}_{\text{base}-(5,1)} \) is
\(
\textbf{H}_{\text{base}-(5,1)} = 
\)
\[
\scriptsize
\setlength{\arraycolsep}{2.5pt} % Adjust column spacing
\centering
\left[\begin{array}{*{25}{c}} % This defines an array with 25 columns
    \R{1} & 0 & 0 & 0 & 0 & 0 & 0 & 0 & 0 & \B{1} & 0 & 0 & 0 & 0 & 0 & 0 & \G{1} & 0 & 0 & 0 & 0 & \PP{1} & 0 & 0 & 1 \\
    \R{1} & \R{1} & 0 & 0 & 0 & 0 & 0 & 0 & 0 & 0 & \B{1} & 0 & 0 & 0 & 0 & 0 & 0 & \G{1} & 0 & 0 & 0 & 0 & \PP{1} & 0 & 0 \\
    0 & \R{1} & \R{1} & 0 & 0 & 0 & 0 & 0 & 0 & 0 & 0 & \B{1} & 0 & 0 & 0 & 0 & 0 & 0 & \G{1} & 0 & 0 & 0 & 0 & \PP{1} & 0 \\
    0 & 0 & \R{1} & \R{1} & 0 & 0 & 0 & 0 & 0 & \B{1} & 0 & 0 & \B{1} & 0 & 0 & 0 & 0 & 0 & 0 & \G{1} & 0 & 0 & 0 & 0 & 0 \\
    0 & 0 & 0 & \R{1} & \R{1} & 0 & 0 & 0 & 0 & 0 & \B{1} & 0 & 0 & \B{1} & 0 & 0 & 0 & 0 & 0 & 0 & \G{1} & 0 & 0 & 0 & 0 \\
    0 & 0 & 0 & 0 & \R{1} & \R{1} & 0 & 0 & 0 & 0 & 0 & \B{1} & 0 & 0 & \B{1} & 0 & \G{1} & 0 & 0 & 0 & 0 & 0 & 0 & 0 & 0 \\
    0 & 0 & 0 & 0 & 0 & \R{1} & \R{1} & 0 & 0 & 0 & 0 & 0 & \B{1} & 0 & 0 & \B{1} & 0 & \G{1} & 0 & 0 & 0 & 0 & 0 & 0 & 0 \\
    0 & 0 & 0 & 0 & 0 & 0 & \R{1} & \R{1} & 0 & 0 & 0 & 0 & 0 & \B{1} & 0 & 0 & 0 & 0 & \G{1} & 0 & 0 & \PP{1} & 0 & 0 & 0 \\
    0 & 0 & 0 & 0 & 0 & 0 & 0 & \R{1} & \R{1} & 0 & 0 & 0 & 0 & 0 & \B{1} & 0 & 0 & 0 & 0 & \G{1} & 0 & 0 & \PP{1} & 0 & 0 \\
    0 & 0 & 0 & 0 & 0 & 0 & 0 & 0 & \R{1} & 0 & 0 & 0 & 0 & 0 & 0 & \B{1} & 0 & 0 & 0 & 0 & \G{1} & 0 & 0 & \PP{1} & 1
\end{array}
\right].
\]
A decagon of matrix $\textbf{H}_{\text{base}-(5,1)}$ for $t=5$ is plotted in Fig. \ref{fig: decagon_loop4}.
\end{example}

Note that the total number of columns is $\sum_{i = 1}^{t}(2i-1) = t^2$, as mentioned before. Fig. \ref{fig: decagon_loop4} shows a decagon corresponding to this base matrix. More specifically, the $i$-th row of the matrix $\textbf{H}_{\text{base-(5,1)}}$ corresponds to the $i$-th vertex of the decagon. Then each column with its two ones defines an edge in the graph, connecting the two corresponding vertices. 
For example, the eleventh column has ones at rows 2 and 5, corresponding to an edge connecting vertices 2 and 5.
We use the same edge color for the decagon as the color of the corresponding entries in the base matrix. Now, consider the \textit{base} code with the matrix $\textbf{H}_{\text{base}-(5,1)}$ as its parity check matrix. Then the vector with ones at positions 1, 2, 3, and 10 is a codeword and corresponds to a cycle of length $4$ with vertices indexed by $1,2,3,4$ in Fig. \ref{fig: decagon_loop4}. This codeword has a weight of 4. In general, it is straightforward to show that the following holds for any $t$.

\begin{remark}
In the corresponding graph representation of \(\mathbf{H}_{\text{base-(t,1)}}\), the number of cycles with length 4 is equal to $A_4$; the number of codewords with weight 4.
\end{remark}

The density of ones in the base parity-check matrix of our FDPC code is given by 
\(\frac{2t^2}{t^2 \times 2t} = \frac{1}{t}\)
which decreases with increasing \(t\). In contrast, for a typical LDPC code, each row contains only a constant number of ones, so the density of ones decreases on the order of \(\mathcal{O}(1/t^2)\), assuming the number of columns scales as \(t^2\). 
We introduce a low-complexity encoding algorithm by modifying the parity-check matrix \(\mathbf{H}_{\text{base-(t,1)}}\) such that the first $2t$ columns contain a square matrix

\begin{equation*}
\scriptsize
    \mathbf{A} = 
    \left[
    \begin{array}{ccccc}
    \R{1} & 0 & 0 & 0 & 0 \\
    \R{1} & \R{1} & 0 & 0 & 0 \\
    0 & \R{1} & \R{1} & 0 & 0  \\
    0 & 0 & \textcolor{red}{\ddots} & \textcolor{red}{\ddots} & 0  \\
    0 & 0 & 0 & \R{1} & \R{1} \\
    \end{array}
    \right].
\end{equation*}
Note that this matrix can be readily obtained from the first \(2t\) columns of \(\mathbf{H}_{\text{base-(t,1)}}\) by modifying only the \(2t\)-th column. Specifically, all elements of the \(2t\)-th column should be set to zero, except for the \(2t\)-th element, which should be set to 1.
To explain the encoding procedure, let us assume that we have the following parity-check matrix.

\(
\textbf{H}_{\text{mbase-(5,1)}} = 
\)
\[
\scriptsize
\setlength{\arraycolsep}{2pt} % Adjust column spacing
\centering
\begin{bNiceArray}[first-row,first-col]{ccccccccccccccccccccccccc}
    & p_1 & p_2 & . & . & . &  &  &  & &  & m_1 & . & . & . &  &  &  &  & & & &  &  &  & \\
    & \R{1} & 0 & 0 & 0 & 0 & 0 & 0 & 0 & 0 & 0 & 0 & 0 & 0 & 0 & 0 & 0 & \G{1} & 0 & 0 & 0 & 0 & \PP{1} & 0 & 0 & 1 \\
    & \R{1} & \R{1} & 0 & 0 & 0 & 0 & 0 & 0 & 0 & 0 & \B{1} & 0 & 0 & 0 & 0 & 0 & 0 & \G{1} & 0 & 0 & 0 & 0 & \PP{1} & 0 & 0 \\
    & 0 & \R{1} & \R{1} & 0 & 0 & 0 & 0 & 0 & 0 & 0 & 0 & \B{1} & 0 & 0 & 0 & 0 & 0 & 0 & \G{1} & 0 & 0 & 0 & 0 & \PP{1} & 0 \\
    & 0 & 0 & \R{1} & \R{1} & 0 & 0 & 0 & 0 & 0 & 0 & 0 & 0 & \B{1} & 0 & 0 & 0 & 0 & 0 & 0 & \G{1} & 0 & 0 & 0 & 0 & 0 \\
    & 0 & 0 & 0 & \R{1} & \R{1} & 0 & 0 & 0 & 0 & 0 & \B{1} & 0 & 0 & \B{1} & 0 & 0 & 0 & 0 & 0 & 0 & \G{1} & 0 & 0 & 0 & 0 \\
    & 0 & 0 & 0 & 0 & \R{1} & \R{1} & 0 & 0 & 0 & 0 & 0 & \B{1} & 0 & 0 & \B{1} & 0 & \G{1} & 0 & 0 & 0 & 0 & 0 & 0 & 0 & 0 \\
    & 0 & 0 & 0 & 0 & 0 & \R{1} & \R{1} & 0 & 0 & 0 & 0 & 0 & \B{1} & 0 & 0 & \B{1} & 0 & \G{1} & 0 & 0 & 0 & 0 & 0 & 0 & 0 \\
    & 0 & 0 & 0 & 0 & 0 & 0 & \R{1} & \R{1} & 0 & 0 & 0 & 0 & 0 & \B{1} & 0 & 0 & 0 & 0 & \G{1} & 0 & 0 & \PP{1} & 0 & 0 & 0 \\
    & 0 & 0 & 0 & 0 & 0 & 0 & 0 & \R{1} & \R{1} & 0 & 0 & 0 & 0 & 0 & \B{1} & 0 & 0 & 0 & 0 & \G{1} & 0 & 0 & \PP{1} & 0 & 0 \\
    & 0 & 0 & 0 & 0 & 0 & 0 & 0 & 0 & \R{1} & \R{1} & 0 & 0 & 0 & 0 & 0 & \B{1} & 0 & 0 & 0 & 0 & \G{1} & 0 & 0 & \PP{1} & 1
\end{bNiceArray}.
\]

This is a full-row rank matrix of size $10$ by $25$ that results in an FDPC$(25, 15)$ code.
The first row contains only the first parity bit, \( p_1 \), which can be obtained as \(p_1 = m_7 + m_{12} + m_{15} \) since the only entry corresponding to \( m_7, m_{12} \) and $m_{15}$ are
1 in the first row. 
After obtaining \( p_1 \), from the second row we can compute the second parity bit as \(p_2 = p_1 + m_1 + m_8 + m_{13}.\) 
Similarly, the remaining parity bits can be obtained sequentially.
In general, the parity check matrix can be modified as

\begin{equation*}
\scriptsize
    \mathbf{H}_{\text{mbase-(t,1)}} = [\mathbf{A} | \mathbf{B}] = 
    \left[
    \begin{array}{ccccc|cccc}
    \R{1} & 0 & 0 & 0 & 0 & \multicolumn{4}{c}{} \\
    \R{1} & \R{1} & 0 & 0 & 0 &  \multicolumn{4}{c}{} \\
    0 & \R{1} & \R{1} & 0 & 0 &  \multicolumn{4}{c}{$\hspace{.5cm} $\mathbf{B}$ \hspace{.5cm}$} \\
    0 & 0 & \textcolor{red}{\ddots} & \textcolor{red}{\ddots} & 0 & \multicolumn{4}{c}{} \\
    0 & 0 & 0 & \R{1} & \R{1} & \multicolumn{4}{c}{} \\
    \end{array}
    \right].
\end{equation*}

% \begin{figure}[htbp] 
% \centering
% 	\includegraphics [width = .8\columnwidth]{./figs/Encoder_Base.png}
% 	\caption{Encoder for the base parity check matrix.} 
% 	\label{fig: encoder_base1}
% \end{figure}

A code constructed based on the parity-check matrix \(\mathbf{H}_{\text{mbase}-(t,1)}\) has a rate of \(\frac{t^2 - 2t}{t^2}\). As \(t \to \infty\), the rate of the code approaches 1. 
We introduce a systematic encoder for this parity-check matrix. For a message \(\mathbf{m} = (m_1, m_2, \dots, m_K)\), the corresponding parity-check vector is denoted as \(\mathbf{p} = (p_1, p_2, \dots, p_{m})\), and the resulting codeword is \(\mathbf{c} = (\mathbf{p}, \mathbf{m})\). The parity bits are computed sequentially: \(p_1\) is derived from the first row, \(p_2\) from the second row, etc. The encoder diagram is shown in Fig. \ref{fig: encoder_base1}, and we use the same encoder for our generalized FDPC code construction (Here, $\mathbf{b}_i$ is the $i$th row of matrix $\mathbf{B}$).

\begin{figure}[htbp] 
\centering
\begin{tikzpicture}
    % Define styles
    \tikzstyle{block} = [draw, rectangle, minimum height=1.5em, minimum width=1.5em]
    
    % Inputs
    \node at (1,0) (input) {$\mathbf{m}$};
    \node at (2,1) (bi) {$\mathbf{b}_i$};
    
    % Dot product (with no gaps, anchored properly)
    \node at (2,0) (dot) [anchor=center, inner sep=-.1pt] {\Large $\odot$};
    \draw[->, line width=.75pt] (input) -- (dot.west);
    \draw[->, line width=.75pt] (bi) -- (dot.north);
    
    % XOR node (with no gaps, anchored properly)
    \node at (3,0) (xor) [anchor=center, inner sep=-.1pt] {\Large $\oplus$};
    \draw[->, line width=.75pt] (dot.east) -- (xor.west);
    
    % Feedback line (from delay to XOR)
    \draw[->, line width=.75pt] (4.75,0) -- ++(0,.75) -- ++(-1.75,0) -- (xor.north);
    
    % XOR to Delay block
    \node[block] (delay) at (4,0) {$D$};
    \draw[->, line width=.75pt] (xor.east) -- (delay.west) node[midway,above] {$p_{i}$};
    
    % Delay to Output
    \draw[->, line width=.75pt] (delay.east) -- ++(1.25,0) node[above] {$p_{i-1}$};
\end{tikzpicture}
\caption{Encoder for the modified base parity check matrix.} 
\label{fig: encoder_base1}
\end{figure}

\subsection{Random Permutations}
As described earlier, for a given $t$, the base parity-check matrix $\textbf{H}_{\text{base}-(t,1)}$ consists of column groups of sizes $2t-1, 2t-3, \cdots, 5, 3, 1$. 
For a given \(t\), blocklength \(N \leq t^2\), number of permutations \texttt{num\_per}, and a given $base$, we construct a parity-check matrix $\mathbf{H}$ for an FDPC\((N, N - 2t \times (\texttt{num\_per}+1))\) code, as described in Algorithm \ref{alg:1}.
The matrix \(\mathbf{H}\) has \(\texttt{m\_size} = 2t \times (\texttt{num\_per}+1)\) rows, and the first \texttt{m\_size} columns exhibit a bidiagonal structure, allowing the FDPC code to be encoded using the encoder shown in Fig. \ref{fig: encoder_base1}, as described in lines 12 to 17 of Algorithm \ref{alg:1}.
For the columns starting from the $(\texttt{m\_size}+1)$-th column of the base matrix, which we refer to as submatrix \(\mathbf{C}\), we perform \texttt{num\_per} random column permutations and append them below submatrix \(\mathbf{C}\), as described in lines 4 to 11. Additionally, in lines 19 to 21, we remove columns from the resulting parity-check matrix \(\mathbf{H}\), starting from column \(\texttt{m\_size} + 1\), until the matrix \(\mathbf{H}\) contains exactly \(N\) columns.
The obtained parity check matrix $\mathbf{H}$ has a format demonstrated as follows:
\begin{equation*}
\scriptsize
    \mathbf{H} = 
    \left[
    \begin{array}{ccccc|c}
    \R{1} & 0 & 0 & 0 & 0 & \multirow{2}{*}{\(\mathbf{C}\)} \\
    \R{1} & \R{1} & 0 & 0 & 0 &  \\
    0 & \R{1} & \R{1} & 0 & 0 & \raisebox{0.1cm}{\(\pi_1(\mathbf{C})\)} \\
    0 & 0 & \textcolor{red}{\ddots} & \textcolor{red}{\ddots} & 0 & \raisebox{0.1cm}{\vdots} \\
    0 & 0 & 0 & \R{1} & \R{1} & \raisebox{0.1cm}{\(\pi_{\texttt{num\_per}}(\mathbf{C})\)} \\
    \end{array}
    \right].
\end{equation*}

\begin{algorithm}[t] 
\footnotesize
\caption{FDPC Code Construction}\label{alg:1}
\begin{algorithmic}[1]
\Statex \hspace*{-\algorithmicindent} \textbf{Input:} $t$, $num\_per$, $N$, \text{base}
\Statex \hspace*{-\algorithmicindent} \textbf{Output:} $\mathbf{H}$ 

\State $m\_base \gets 2t$

\State $m\_size \gets m\_base + num\_per \times m\_base$

\State $\mathbf{H}, n \gets \text{generate\_base\_matrix}(t, base)$ % base=1--> n = 2^2, base=2 --> n=t(t+1)/2

\State $\mathbf{C} \gets \mathbf{H}[:, m\_size+1:\text{end}]$

\For {$i = 1 \textbf{ to } num\_per$}
    \State $col\_permutation \gets \text{randperm}(\text{size}(\mathbf{C}, 2))$
    \State $\mathbf{C}\_permuted \gets \mathbf{C}[:, col\_permutation]$ 
    \State $\mathbf{H}\_permuted \gets zeros(m\_base, n)$
    \State $\mathbf{H}\_permuted[:, m\_size+1:\text{end}] \gets \mathbf{C}\_permuted$
    \State $\mathbf{H} \gets [\mathbf{H}; \mathbf{H}\_permuted]$
\EndFor

\State \Comment{For the encoder}
\State $\mathbf{H}[1:m\_size, 1:m\_size] \gets \mathbf{0}$
\For {$i = 1 \textbf{ to } m\_size-1$}
    \State $\mathbf{H}[i,i] \gets 1$
    \State $\mathbf{H}[i+1,i] \gets 1$
\EndFor
\State $\mathbf{H}[m\_size, m\_size] \gets 1$

\If {$N \neq n$}
    \State Remove columns \( m\_size+1 \) to \( m\_size + (n - N) \) from \(\mathbf{H}\)
\EndIf

\end{algorithmic}
\end{algorithm}

\section{FDPC Code Construction base-II}\label{sec:base2}

In order to generalize the FDPC code construction, one possible way is to have a graph representation with a minimum cycle of length $d$, where $d$ is an even number. Then this corresponds to a base code with a minimum distance of $d$. To this end, we design a matrix \( \textbf{H}_{\text{base}-(t,2)} \) of size \( (2t, \frac{t(t+1)}{2}) \) for $d=6$. 
Each column contains exactly two ones. The first \( 2t-1 \) columns have two ones with no zeros between them. 
The next \( 2t-5 \) columns have four zeros between the ones. 
The next \( 2t-9 \) columns have eight zeros between the ones, etc. 
The number of columns of the base matrix \( \textbf{H}_{\text{base-(t,2)}} \) for an odd $t$ can be obtained as
\begin{equation*}\label{eq: colNum2}
\begin{split}
    &2t-1 + 2t-5 + \cdots + 5 + 1= \sum_{i = 1}^{\frac{t+1}{2}}(4i-3) =  \frac{t(t+1)}{2}.\\
\end{split}
\end{equation*}
For an even $t$, the number of columns can be obtained as
\begin{equation*}
\begin{split}
    &2t-1 + 2t-5 + \cdots + 7 + 3 = \sum_{i = 1}^{\frac{t}{2}}(4i-1) = \frac{t(t+1)}{2}.
\end{split}
\end{equation*}

\begin{example}
For $t=5$, the corresponding base matrix is
\[
\scriptsize
\setlength{\arraycolsep}{3pt} % Adjust column spacing
\centering
\textbf{H}_{\text{base-(5,2)}} = \left[\begin{array}{*{15}{c}} % This defines an array with 25 columns
    \R{1} & 0 & 0 & 0 & 0 & 0 & 0 & 0 & 0 & \G{1} & 0 & 0 & 0 & 0 & 1 \\
    \R{1} & \R{1} & 0 & 0 & 0 & 0 & 0 & 0 & 0 & 0 & \G{1} & 0 & 0 & 0 & 0 \\
    0 & \R{1} & \R{1} & 0 & 0 & 0 & 0 & 0 & 0 & 0 & 0 & \G{1} & 0 & 0 & 0 \\
    0 & 0 & \R{1} & \R{1} & 0 & 0 & 0 & 0 & 0 & 0 & 0 & 0 & \G{1} & 0 & 0 \\
    0 & 0 & 0 & \R{1} & \R{1} & 0 & 0 & 0 & 0 & 0 & 0 & 0 & 0 & \G{1} & 0 \\
    0 & 0 & 0 & 0 & \R{1} & \R{1} & 0 & 0 & 0 & \G{1} & 0 & 0 & 0 & 0 & 0 \\
    0 & 0 & 0 & 0 & 0 & \R{1} & \R{1} & 0 & 0 & 0 & \G{1} & 0 & 0 & 0 & 0 \\
    0 & 0 & 0 & 0 & 0 & 0 & \R{1} & \R{1} & 0 & 0 & 0 & \G{1} & 0 & 0 & 0 \\
    0 & 0 & 0 & 0 & 0 & 0 & 0 & \R{1} & \R{1} & 0 & 0 & 0 & \G{1} & 0 & 0 \\
    0 & 0 & 0 & 0 & 0 & 0 & 0 & 0 & \R{1} & 0 & 0 & 0 & 0 & \G{1} & 1
\end{array}
\right].
\]
The corresponding decagon of matrix $\textbf{H}_{\text{base}-(5,2)}$ for $t=5$ and minimum cycle of length 6 is plotted in Fig. \ref{fig: decagon_loop6}.

\end{example}

For a given $t$ and $base$, Algorithm \ref{alg:2} generates a base parity-check matrix.
The variable \texttt{gap} is initialized to $0$ in line 3 of Algorithm~\ref{alg:2}. This variable determines the gap between the two ones in each column, as described in lines 9 and 10 of the algorithm. The gap increases by 2 for each group of columns in line 13. Additionally, this variable determines the size of each group of columns, which is used to run the iterations over the diagonals of the group submatrices in line 5. 

The variable \texttt{col} tracks the current column of the base matrix and increments until it reaches $t^2$. Lines 4 to 13 construct the \( \mathbf{H}_{\text{base}-(t,1)} \) matrix. 

For $base = 2$, we remove the columns that have 2, 6, 10, $\cdots$ zeros between the ones from the constructed \( \mathbf{H}_{\text{base}-(t,1)} \) matrix in lines 15 to 28. The variable \texttt{current\_gap} determines the number of zeros between the ones corresponding to the columns that need to be removed. It is initialized to $0$ and increases by $4$ in each iteration. The variable \texttt{valid\_columns} stores the indices of the columns that need to be kept, initially containing all indices from 1 to $t^2$. We remove the invalid columns at line 22; these are the columns where the distance between the ones is $\texttt{current\_gap} + 1$. 
For example, when $\texttt{current\_gap} = 2$, the distance between the ones is 3, and these columns are removed.

\begin{algorithm}
\footnotesize
\caption{Base Matrix}\label{alg:2}
\begin{algorithmic}[1]
\Statex \hspace*{-\algorithmicindent} \textbf{Input:} $t$, $base$
\Statex \hspace*{-\algorithmicindent} \textbf{Output:} $\mathbf{H}$, $\text{size}(\mathbf{H}, 2)$ 

\State $\mathbf{H} \gets \mathbf{0}$ \Comment{Initialize $\mathbf{H}$ of size $2t$ by $t^2$.}
\State $col \gets 1$ \Comment{Initialize the column index}
\State $gap \gets 0$ \Comment{No gap between ones at the start}

\While {$col \leq t^2$}
    \For {$row = 1$ \textbf{to} $2t - gap - 1$}
        \If {$col > t^2$}
            \State \textbf{break} \Comment{Break if the column index exceeds $t^2$}
        \EndIf
        \State $\mathbf{H}[row, col] \gets 1$ \Comment{Place 1 in the current row and column}
        \State $\mathbf{H}[row + gap + 1, col] \gets 1$ \Comment{Place 1 with a gap in the same column}
        \State $col \gets col + 1$ \Comment{Move to the next column}
    \EndFor
    \State $gap \gets gap + 2$ \Comment{Increase the gap by 2 for the next group of columns}
\EndWhile

\If {$base == 2$} 
    \State \Comment{Step 1: Remove columns where the gap is 2, 6, etc.}
    \State $valid\_columns \gets \mathbf{true}(1, \text{size}(\mathbf{H}, 2))$ \Comment{All col initially valid}
    \State $current\_gap \gets 2$ \Comment{Start from a gap of 2 zeros between ones}
    
    \While {$current\_gap < \text{size}(\mathbf{H}, 1) - 1$}
        \For {$col = 1$ \textbf{to} $\text{size}(\mathbf{H}, 2)$}
            \If {\text{any(diff(find(H(:, col) == 1)) == current\_gap + 1)}}
                \State $valid\_columns[col] \gets \mathbf{false}$ \Comment{Mark column for removal}
            \EndIf
        \EndFor
        \State $current\_gap \gets current\_gap + 4$ \Comment{2, 6, etc.}
    \EndWhile
    
    \State $\mathbf{H} \gets \mathbf{H}(:, valid\_columns)$ \Comment{Keep only the valid columns}
\EndIf

\end{algorithmic}
\end{algorithm}

\begin{remark}
    For an odd $ t = 2k + 1 $, the base matrix $ \mathbf{H}_{\text{base-(t,2)}} $ is a regular matrix with row weights of $ \frac{t+1}{2} $. However, for an even $ t = 2k $, the base matrix has row weights of $ \lceil \frac{t}{2} \rceil $, except for the first and last rows, which have weights of $ \lfloor \frac{t}{2} \rfloor $.
\end{remark}

\section{Numerical results}\label{sec:numerical}
In this section, we demonstrate simulations over the binary-input additive white Gaussian noise (BI-AWGN) channel model with a binary phase-shift keying (BPSK) modulation. For the FDPC and LDPC codes, we employ the normalized min-sum algorithm.
For the polar codes, we employ the scaled min-sum algorithm with a scaling factor of $0.9375$ \cite{yuan2014early}.
The ECP metrics are the frame error rate (FER) and bit error rate (BER).

\subsection{FDPC(256, 192) code}
\begin{figure}[htbp] 
\centering
	\includegraphics [width = \columnwidth]{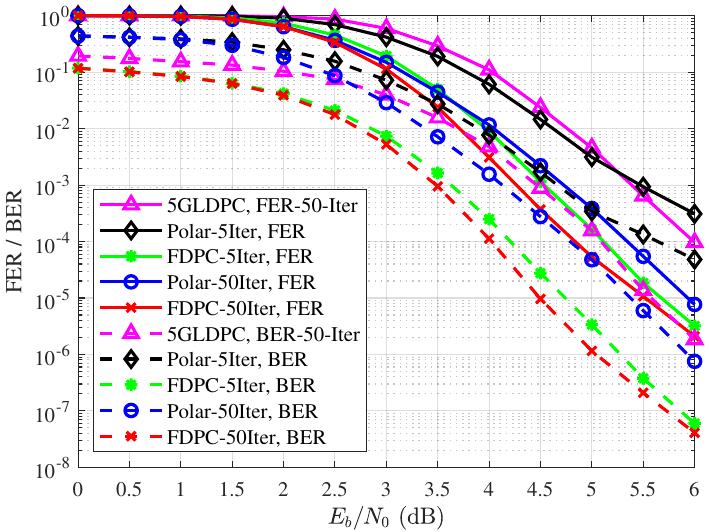}
	\caption{ECP comparison of $(256,192)$ FDPC, 5G LDPC, and polar codes.} 
	\label{fig: polarBP_FDPCminsum_N256_K192}
\end{figure}

Starting from base~1 with a single permutation, we construct a parity-check matrix of size $64 \times 256$ with $t = 16$, resulting in an FDPC$(256, 192)$ code. 
Figure~\ref{fig: polarBP_FDPCminsum_N256_K192} shows the FER and BER performance of this code using the normalized min-sum algorithm with both 50 and 5 decoding iterations. 
For comparison, the performance of the polar$(256, 192)$ code—constructed using Gaussian approximation at $E_b/N_0 = 4\,\mathrm{dB}$—is also plotted in the same figure under BP decoding with 50 and 5 iterations. 
Additionally, the performance of the 5G LDPC code is included for reference. 
Notably, the FDPC code with just 5 iterations outperforms the polar code with 50 iterations and achieves approximately a 1~dB coding gain over the polar code with 5 iterations and the 5G LDPC code with 50 iterations at an FER level of $10^{-3}$.\footnote{All of the codes simulated in this paper are available at \href{https://github.com/moradi-coding/}{https://github.com/moradi-coding/}.}

\subsection{FDPC(128, 80) code}
\begin{figure}[htbp] 
\centering
	\includegraphics [width = \columnwidth]{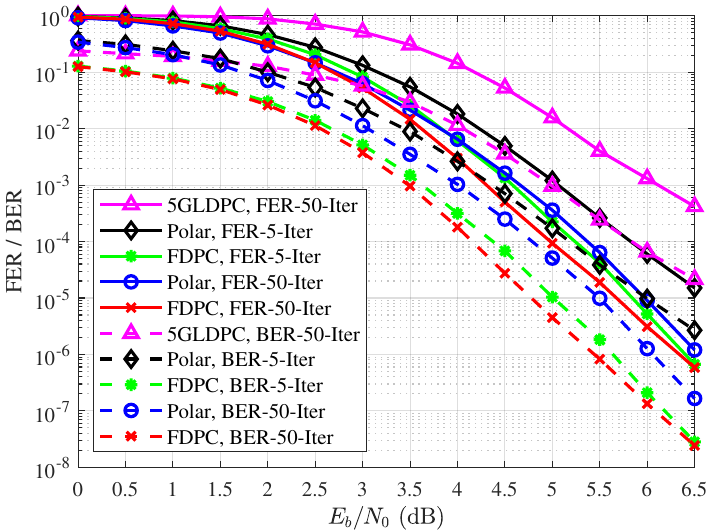}
	\caption{ECP comparison of $(128,80)$ FDPC, 5G LDPC, and polar codes.} 
	\label{fig: polarBP_FDPCminsum_N128_K80}
\end{figure}

Starting from base 1 with one permutation, we first construct a parity-check matrix of size $48 \times 144$ with $t = 12$. 
Then, we remove columns starting from the $(4t + 1) = 49$-th column through the 65th column, resulting in a parity-check matrix of size $48 \times 128$. 
This produces an FDPC$(128, 80)$ code.
Fig. \ref{fig: polarBP_FDPCminsum_N128_K80} shows the FER and BER performance of this FDPC code using the normalized min-sum algorithm with 5 and 50 iterations. 
For comparison, the performance of a polar$(128, 80)$ code (the codeword length of polar codes is a power of two), constructed using the Gaussian approximation and decoded with the BP algorithm under the same number of iterations, is also plotted.
Additionally, the performance of the 5G LDPC code is included for reference. 
Notably, the FDPC code with just 5 iterations outperforms the polar code with 50 iterations and achieves approximately a $0.5$~dB coding gain over the polar code with 5 iterations and a $1.5$~dB coding gain over the 5G LDPC code with 50 iterations at a FER level of $10^{-3}$.

\subsection{FDPC(256, 164) code}

\begin{figure}[htbp] 
\centering
	\includegraphics [width = \columnwidth]{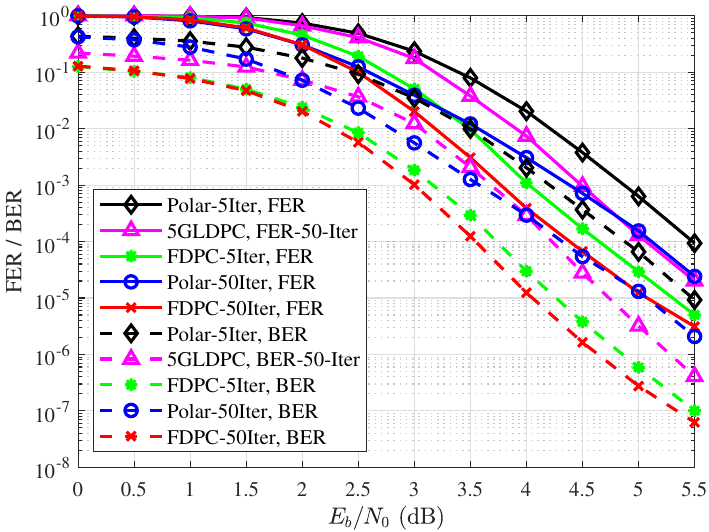}
	\caption{ECP comparison of $(256,164)$ FDPC, 5G LDPC, and polar codes.} 
	\label{fig: polarBP_FDPCminsum_N256_K164}
\end{figure}

Using the $\mathbf{H}_{\text{base-(23,2)}}$ with one permutation, we first construct a $92\times 276$ parity-check matrix.  
By removing columns starting from the 93rd column, we obtain an FDPC$(256, 164)$ code.  
The ECP of this code is plotted in Fig. \ref{fig: polarBP_FDPCminsum_N256_K164} using 5 and 50 iterations of the normalized min-sum algorithm.  
For comparison, the figure also includes the performance of a polar$(256, 164)$ code using BP decoding.  
The polar code is constructed using the Gaussian approximation at $E_b/N_0 = 4~$dB and decoded with the same number of iterations.
Additionally, the performance of the 5G LDPC code is included for reference. 
Notably, the FDPC code with just 5 iterations achieves approximately a $0.5$~dB coding gain over the polar code and 5G LDPC code with 50 iterations at an FER level of $10^{-3}$.

\subsection{FDPC(1024, 844) code}
\begin{figure}[htbp] 
\centering
	\includegraphics [width = \columnwidth]{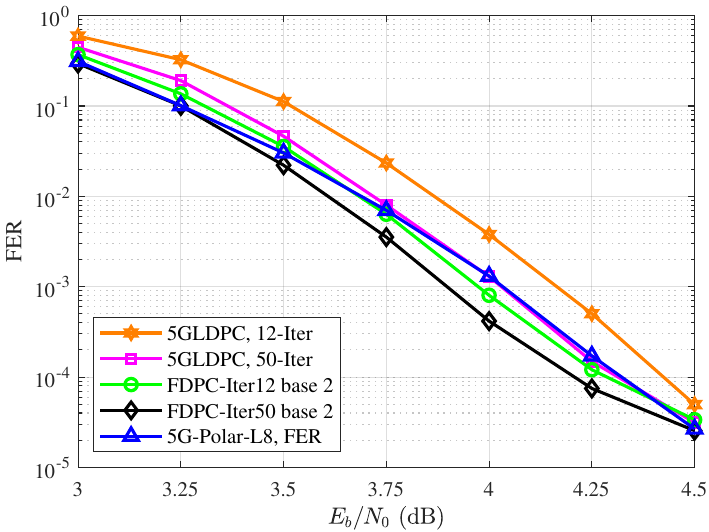}
	\caption{FER performance comparison of $(1024, 844)$ FDPC, 5G-LDPC, and CA-SCL 5G-polar codes.} 
	\label{fig: FER_N1024_K832_K844}
\end{figure}

Using $\mathbf{H}_{\text{base-(45,2)}}$ with one permutation, we first construct a $180 \times 1035$ base matrix. 
By removing columns starting from the 180th column, we obtain an FDPC$(1024, 844)$ code. 
% Similarly, using $\mathbf{H}_{\text{base-(32,1)}}$ with two permutations, we construct a $192 \times 1024$ parity-check matrix to obtain an FDPC$(1024, 832)$ code.
The FER performance of this code is shown in Fig. \ref{fig: FER_N1024_K832_K844} for different numbers of iterations of the normalized min-sum algorithm.
For comparison, we also plot the performance of CA-SCL decoding of 5G polar codes with a CRC length of 24 and a list size of 8.
Additionally, the performance of the 5G LDPC code is included for reference. 
Our proposed FDPC code, with only 12 decoding iterations, achieves an FER performance comparable to that of the CA-SCL decoding of 5G polar codes and the normalized min-sum decoding of 5G LDPC codes with 50 iterations.

\subsection{FDPC(1024, 832) code}
\begin{figure}[htbp] 
\centering
	\includegraphics [width = \columnwidth]{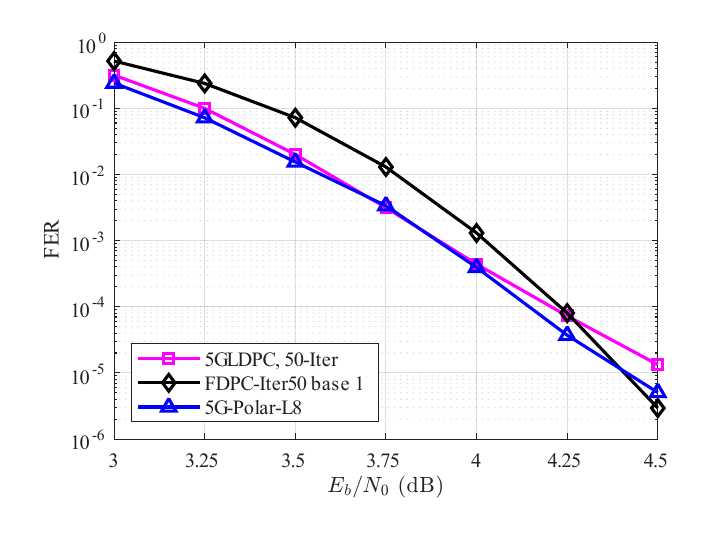}
	\caption{FER performance comparison of $(1024, 832)$ FDPC, 5G-LDPC, and CA-SCL 5G-polar codes.} 
	\label{fig: FDPCminsum_N1024_K832_base1}
\end{figure}

Using $\mathbf{H}_{\text{base-(32,1)}}$ with two permutations, we construct a $192 \times 1024$ parity-check matrix to obtain an FDPC$(1024, 832)$ code.
The FER performance of this code is shown in Fig. \ref{fig: FDPCminsum_N1024_K832_base1} for $50$ iterations of the normalized min-sum algorithm.
Additionally, the performance of the 5G LDPC code is included for reference. 
For comparison, we also plot the performance of CA-SCL decoding of 5G polar codes with a CRC length of 24 and a list size of 8.
At \(E_b/N_0 = 4.5~\text{dB}\), our proposed FDPC code achieves a better ECP compared to both 5G LDPC and polar codes.

\subsection{FDPC(16384, 15660) and FDPC(16384, 15616) codes}

\begin{figure}[htbp] 
\centering
	\includegraphics [width = \columnwidth]{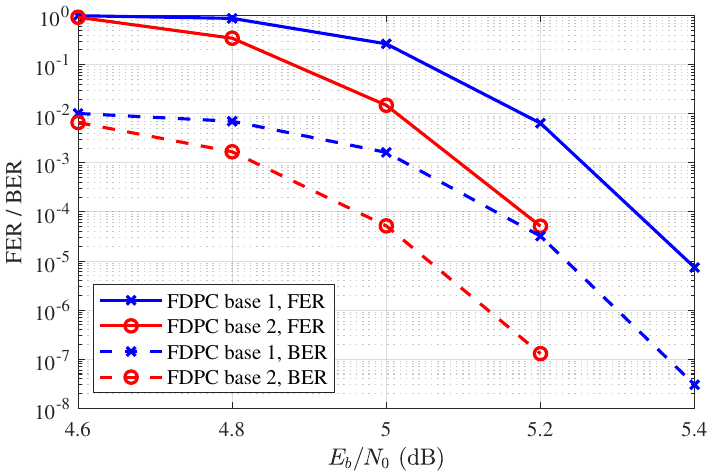}
	\caption{ECP comparison of FDPC$(16384, 15660)$ and FDPC$(16384, 15616)$ codes.} 
	\label{fig: FDPCminsum_N16384_K15660_K15616}
\end{figure}

Using $\mathbf{H}_{\text{base-(181,2)}}$ with one permutation, we first construct a $362 \times 16471$ parity-check matrix. By removing columns starting from the 725th column, we obtain an FDPC$(16384, 15661)$ code. Similarly, using $\mathbf{H}_{\text{base-(128,1)}}$ with two permutations, we construct a $768 \times 16384$ parity-check matrix to obtain an FDPC$(16384, 15616)$ code. 
The ECP of these codes is shown in Fig. \ref{fig: FDPCminsum_N16384_K15660_K15616} using 12 iterations of the normalized min-sum algorithm.

%%%%%%%%%%%%%%%%%%%%%%%%%%%%%%%%%%%%%%%%%%%%%%%%%%%%%%%%%%%%%%%%%%%%%%%%%%%%%%%%%%%%%%%%%%%%%%%%%%%%%%%
\section{Conclusion}\label{sec:Conclusion}
In this paper, we introduced a novel construction and a low-complexity encoder for FDPC codes. At short block lengths, simulation results demonstrate both improved error-correction performance and reduced latency compared to the message passing decoding of polar and 5G LDPC codes. For moderate block lengths, we also compared the performance of our proposed FDPC codes with that of 5G LDPC and 5G polar codes, highlighting the advantages of the proposed design.

%%%%%%%%%%%%%%%%%%%%%%%%%%%%%%%%%%%%%%%%%%%%%%%%%%%%%%%%%%%%%%%%%%%%%%%%%%%%%%%%%%%%%%%%%%%%%%%%%%%%%%%
% \section*{Acknowledgment}

\ifCLASSOPTIONcaptionsoff
  \newpage
\fi

\bibliographystyle{IEEEtran}
\bibliography{bibliography}

\begin{thebibliography}{10}
\providecommand{\url}[1]{#1}
\csname url@samestyle\endcsname
\providecommand{\newblock}{\relax}
\providecommand{\bibinfo}[2]{#2}
\providecommand{\BIBentrySTDinterwordspacing}{\spaceskip=0pt\relax}
\providecommand{\BIBentryALTinterwordstretchfactor}{4}
\providecommand{\BIBentryALTinterwordspacing}{\spaceskip=\fontdimen2\font plus
\BIBentryALTinterwordstretchfactor\fontdimen3\font minus \fontdimen4\font\relax}
\providecommand{\BIBforeignlanguage}[2]{{%
\expandafter\ifx\csname l@#1\endcsname\relax
\typeout{** WARNING: IEEEtran.bst: No hyphenation pattern has been}%
\typeout{** loaded for the language `#1'. Using the pattern for}%
\typeout{** the default language instead.}%
\else
\language=\csname l@#1\endcsname
\fi
#2}}
\providecommand{\BIBdecl}{\relax}
\BIBdecl

\bibitem{gallager1962low}
R.~Gallager, ``Low-density parity-check codes,'' \emph{IRE Transactions on Information Theory}, vol.~8, no.~1, pp. 21--28, 1962.

\bibitem{mackay1995good}
D.~J. MacKay and R.~M. Neal, ``Good codes based on very sparse matrices,'' in \emph{IMA International Conference on Cryptography and Coding}.\hskip 1em plus 0.5em minus 0.4em\relax Springer, 1995, pp. 100--111.

\bibitem{3GPP_2018}
\BIBentryALTinterwordspacing
3GPP, ``{NR}; multiplexing and channel coding,'' Tech. Rep. TS 38.212, Rel. 15, June 2018. [Online]. Available: \url{http://www.3gpp.org/DynaReport/38-series.htm}
\BIBentrySTDinterwordspacing

\bibitem{wang2023road}
C.-X. Wang \emph{et~al.}, ``On the road to {6G}: Visions, requirements, key technologies, and testbeds,'' \emph{IEEE Communications Surveys \& Tutorials}, vol.~25, no.~2, pp. 905--974, 2023.

\bibitem{mitchell2015spatially}
D.~G.~M. Mitchell, M.~Lentmaier, and D.~J. Costello, ``Spatially coupled {LDPC} codes constructed from protographs,'' \emph{IEEE Transactions on Information Theory}, vol.~61, no.~9, pp. 4866--4889, 2015.

\bibitem{tanner2004ldpc}
R.~Tanner, D.~Sridhara, A.~Sridharan, T.~Fuja, and D.~Costello, ``{LDPC} block and convolutional codes based on circulant matrices,'' \emph{IEEE Transactions on Information Theory}, vol.~50, no.~12, pp. 2966--2984, 2004.

\bibitem{pusane2011deriving}
A.~E. Pusane, R.~Smarandache, P.~O. Vontobel, and D.~J. Costello, ``Deriving good {LDPC} convolutional codes from ldpc block codes,'' \emph{IEEE Transactions on Information Theory}, vol.~57, no.~2, pp. 835--857, 2011.

\bibitem{mitchell2014quasi}
D.~G.~M. Mitchell, R.~Smarandache, and D.~J. Costello, ``Quasi-cyclic {LDPC} codes based on pre-lifted protographs,'' \emph{IEEE Transactions on Information Theory}, vol.~60, no.~10, pp. 5856--5874, 2014.

\bibitem{costello2014spatially}
D.~J. Costello, L.~Dolecek, T.~E. Fuja, J.~Kliewer, D.~G. Mitchell, and R.~Smarandache, ``Spatially coupled sparse codes on graphs: theory and practice,'' \emph{IEEE Communications Magazine}, vol.~52, no.~7, pp. 168--176, 2014.

\bibitem{bocharova2022design}
I.~E. Bocharova \emph{et~al.}, ``Design and analysis of {NB} {QC-LDPC} codes over small alphabets,'' \emph{IEEE Transactions on Communications}, vol.~70, no.~5, pp. 2964--2976, 2022.

\bibitem{arikan2009channel}
E.~Ar{\i}kan, ``Channel polarization: A method for constructing capacity-achieving codes for symmetric binary-input memoryless channels,'' \emph{IEEE Transactions on Information Theory}, vol.~55, no.~7, pp. 3051--3073, 2009.

\bibitem{tal2015list}
I.~Tal and A.~Vardy, ``List decoding of polar codes,'' \emph{IEEE Transactions on Information Theory}, vol.~61, no.~5, pp. 2213--2226, 2015.

\bibitem{mahdavifar2024high}
\BIBentryALTinterwordspacing
H.~Mahdavifar, ``High-rate fair-density parity-check codes,'' 2024. [Online]. Available: \url{https://arxiv.org/abs/2402.06814}
\BIBentrySTDinterwordspacing

\bibitem{richardson2018design}
T.~Richardson and S.~Kudekar, ``Design of low-density parity check codes for {5G} new radio,'' \emph{IEEE Communications Magazine}, vol.~56, no.~3, pp. 28--34, 2018.

\bibitem{moradi2024polarization}
\BIBentryALTinterwordspacing
M.~Moradi, ``Polarization-adjusted convolutional ({PAC}) codes as a concatenation of inner cyclic and outer polar- and {R}eed-{M}uller-like codes,'' \emph{Finite Fields and Their Applications}, vol.~93, p. 102321, 2024. [Online]. Available: \url{https://www.sciencedirect.com/science/article/pii/S1071579723001636}
\BIBentrySTDinterwordspacing

\bibitem{arikan2010polar}
E.~Ar{\i}kan, ``Polar codes: A pipelined implementation,'' in \emph{Proc. 4th ISBC}, vol. 2010, 2010, pp. 11--14.

\bibitem{yuan2014early}
B.~Yuan and K.~K. Parhi, ``Early stopping criteria for energy-efficient low-latency belief-propagation polar code decoders,'' \emph{IEEE Transactions on Signal Processing}, vol.~62, no.~24, pp. 6496--6506, 2014.

\end{thebibliography}

\end{document}